\documentclass[journal]{IEEEtran}

\usepackage{bm}

\usepackage{diagbox}
\usepackage{multirow}

\usepackage{graphicx}
\graphicspath{{./picture/}}

\usepackage{amsmath}

\usepackage{enumerate}

\usepackage{url}

\usepackage{cite}

\usepackage{color}

\begin{document}

\title{A Deep Learning Reconstruction Framework for Differential Phase-Contrast Computed Tomography with Incomplete Data}

\author{Jianbing~Dong
        and~Jian~Fu
        and~Zhao~He 
\thanks{Jian Fu, Jianbing Dong and Zhao he are with Research Center of Digital Radiation Imaging, Beijing University of Aeronautics and Astronautics, 100191 Beijing, China. Email: fujian706@buaa.edu.cn.}
}


\maketitle

\begin{abstract}
Differential phase-contrast computed tomography (DPC-CT) is a powerful analysis tool for soft-tissue and low-atomic-number samples. Limited by the implementation conditions, DPC-CT with incomplete projections happens quite often. Conventional reconstruction algorithms are not easy to deal with incomplete data. They are usually involved with complicated parameter selection operations, also sensitive to noise and time-consuming. In this paper, we reported a new deep learning reconstruction framework for incomplete data DPC-CT. It is the tight coupling of the deep learning neural network and DPC-CT reconstruction algorithm in the phase-contrast projection sinogram domain. The estimated result is the complete phase-contrast projection sinogram not the artifacts caused by the incomplete data. After training, this framework is determined and can reconstruct the final DPC-CT images for a given incomplete phase-contrast projection sinogram. Taking the sparse-view DPC-CT as an example, this framework has been validated and demonstrated with synthetic and experimental data sets. Embedded with DPC-CT reconstruction, this framework naturally encapsulates the physical imaging model of DPC-CT systems and is easy to be extended to deal with other challengs. This work is helpful to push the application of the state-of-the-art deep learning theory in the field of DPC-CT.
\end{abstract}

\section{Introduction}

\IEEEPARstart{T}{he} invention of X-ray computed tomography (CT) has led to a revolution in many fields such as medical imaging, nondestructive testing and materials science. However, since traditional X-ray contrast is generated by the difference in attenuation, weak-absorbing materials are not imaged satisfactorily, limiting the range of application of X-ray CT. To overcome this problem, X-ray phase-contrast CT (PC-CT) uses the phase shift that X-rays undergo when passing through matter as the imaging signal to nondestructively provide the internal physical and biomedical properties of the specimens. Over the last years, several phase-contrast imaging techniques have been developed \cite{phase1,phase2,phase3,phase4,phase5,phase6,phase7,phase8,phase9,phase10,phase11,phase12,phase13,phase14,phase15,phase16,phase17,phase18,phase19,phase21,phase20,phase22,phase23,phase24}. One of the recent developments is differential PC-CT (DPC-CT), based on a grating interferometer \cite{phase12,phase13,phase14,phase15,phase16,phase17,phase18,phase19,phase21}. It has become more and more popular as a powerful analysis tool for soft-tissue and low-atomic-number samples.

Image reconstruction plays an important role for the development of DPC-CT. Filtered back-projection (FBP) with an imaginary Hilbert filter is generally preferred since it keeps a good balance between reconstrucion speed and image quality when applied to complete data. However, constrained by implementation conditions, DPC-CT with incomplete projections occurs quite frequently. The corresponding FBP reconstruction will have quite visible artifacts and noise.

Reconstruction with incomplete data has attracted more and more interests. Guna Kim et al investigated analytic CT reconstruction in sparse-angular sampling using a new interpolation method \cite{analyticInterpo} to reduce patient radiation dosage. According to the results, the quality of the images reconstructed by their method was considerably improved over the cases using cubic interpolation method. Deriving a maximum likelihood (ML) reconstruction algorithm with regularization \cite{dpcct1} for differential phase-contrast imaging, Thomas et al used spherically symmetric basis functions and differential footprints in forward and back-projection to avoid the need for numerical differentiation. Their results showed that sparsely sampled data could be handled efficiently. Based on a novel spling-based discretization of the forward model and an iterative reconstruction algorithm using the alternating direction method of multipliers, Masih et al gave out an iterative reconstruction method \cite{dpcct2} for DPC-CT with fewer angular views, and the results suggested that their method allows to reduce the number of required views by a factor of four. Jian Fu et al developed an algebraic iteration reconstruction technique \cite{dpcct3} for incomplete data DPC-CT. By minimizing the image total variation, their work could permits accurate tomographic imaging with less data. These reconstruction techniques could be better than FBP, but they still have some limits such as expensive time consumption for the succesive iterative steps and the complicated parameter selection.

A more recent trend is the application of deep learning (DL). It has led to a series of breakthroughs for image classification\cite{classification1,classification2} and segmentation\cite{U-net} and also demonstrated impressive results on signal denoising\cite{image-denoising} and artifacts reduction\cite{reduction1,reduction2}.

There is currently a scarcity of researches on applying DL to DPC-CT reconstruction, and there remains an important need to develop the relative techniques. Nevertheless, several works have applied DL to absorption-based incomplete CT reconstruction. They can be grouped into two categories. The first category can be classified as post-processing. Its key idea is to reduce artifacts on the CT image domain. A method presented by Cierniak \cite{CierniakDlRecon} uses hopfield-type neural network on the back-projection to solve the image deburring problem. This approach bypasses the problem of too many parameters by fixing them in the back-projection step and degenerates the reconstruction to an image-based filtering approach. Based on a persistent homology analysis, Han et al developed a deep learning residual architecture \cite{DLinCT1} for sparse-view CT reconstruction. The input of this architecture is the initial corrupted reconstruction image from FBP or other algorithm. It firstly estimates topologically simpler streaking artifacts from the input image and then subtracts the estimated result from the input image to get artifact-free image. Obviously this method is independent on X-CT reconstruction and works in an indirect way. Using multi-scale wavelet, they extended their work to limited angle CT reconstruction \cite{DLinCT2}. Jin et al also proposed a deep convolutional neural network \cite{convolutionalforCT} for inverse problem in imaging. It is similarly independent on X-CT reconstruction, but the estimated result is the final CT image not the artifacts. With dialted convolutions, Pelt et al introduced an architecture \cite{PeltDlRecon} to capture features at different image scales and densely connect all feature maps with each other. Their method is also independent on CT reconstruction, but is able to achieve accurate results with fewer parameters, which reduced the risk of overfitting the training data. Zhicheng Zhang et al proposed a method \cite{ZhichengDLRecon}, which takes full advantages of DenseNet \cite{denseNet} and deconvolution \cite{deconvolution} to remove streaking artifacts from the sparse-view CT images. The second category can be classified as sinogram completion. Its key idea is to complement the incomplete projections before reconstructing them with analytical algorithms. Hoyeon Lee et al developed an interpolation method \cite{sinoInterpo1} using convolutional neural network (CNN) to interpolate the missing data in sinogram from sparse-view CT. Their work shows better result than other interpolation methods, like linear interpolation method. Changed the network to U-Net \cite{U-net} and combined with residual learning \cite{residuallearning}, their extended work \cite{sinoInterpo2} outperformed the  existing interpolation methods and iterative image reconstruction approach. Donghoong Lee et al employed DL on hybrid domain of sparsely sampled CT to restore high quality images \cite{sinoInterpo3}. Firstly, they apply DL on the linear interpolated sinogram to get full sampled sinogram, then, DL is utilized again on the CT image reconstructed from the full sampled sinogram to obtain the final CT image. Their experiments shows it is able to restore images to a quality similar to fully sampled images.

In this paper, we report a new deep learning reconstruction framework for DPC-CT with incomplete projections. Different from post-processing and sinogram completion methods, we firstly project the initial reconstructed DPC-CT image into corrupted sinogram, in which the missing information will be completed in a fashion of CT scanning instead of interpolation, and then remove the artifacts on the projection sinogram domain. It is the tight coupling of the DL and FBP algorithm in the phase-contrast projection sinogram domain. The estimated result is the complete projection sinogram not the DPC-CT image or the artifacts. Embedded with DPC-CT reconstruction, it naturally encapsulates the physical imaging model of DPC-CT systems and is easy to be extended to deal with other challengs such as photon starvation and phase wrapping. When training, this framework firstly obtains the forward projections from the initial reconstruction by applying FBP to the original incomplete projection sinogram. Taking the complete projection sinogram as a target, they are then fed into the neural network to get the net parameters by deep learning. After training, this framework is determined and can reconstruct the final DPC-CT image for a given incomplete phase-contrast projection sinogram. Taking sparse-view DPC-CT reconstruction as an application example, this framework has been validated by using synthetic data sets and experimental data sets. This work is helpful to push the application of deep learning in the field of DPC-CT.

\section{Methods}

\subsection{Framework overview}

\begin{figure*}[!htbp]
\centering

\includegraphics[width=0.8\textwidth]{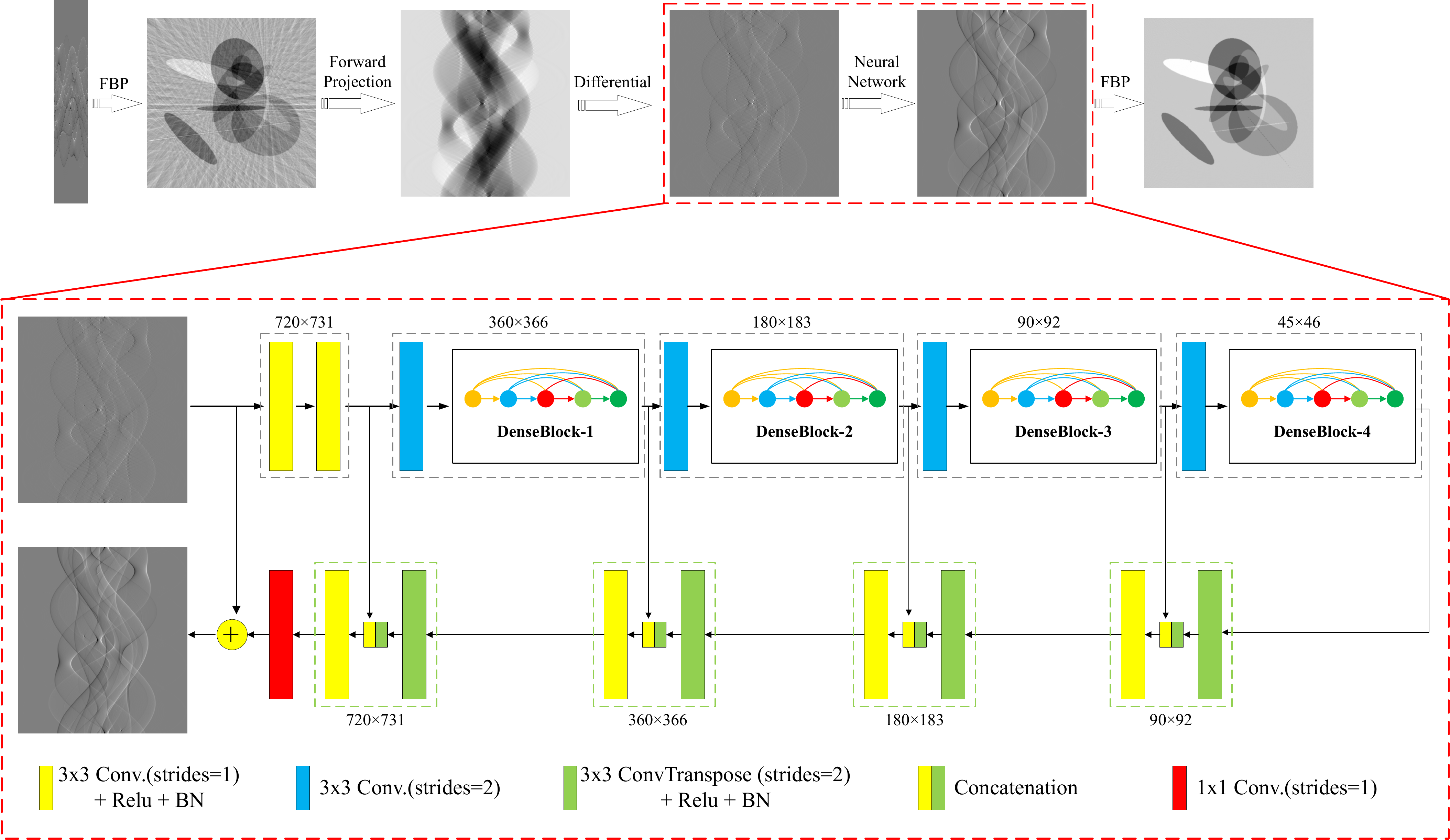}

\caption{The architecture of the framework: DLFBP. The neural network is constructed using U-net and DenseBlock.}

\label{framework}

\end{figure*}

Fig.\ref{framework} shows the proposed deep learning reconstruction framework for DPC-CT with incomplete projections. It is based on FBP and a neural network and called DLFBP. This framework consists of five parts. Initial FBP reconstruction of the original incomplete phase-contrast sinogram is the first part. The second part is the forward projection operator which is applied to the initial reconstructed image to obtain corrupted sinogram. Then the differential operation is conducted to get differential phase-contrast sinogram. The fourth part is a neural network and used to execute deep learning. Final FBP reconstruction of the complete phase-contrast sinogram from the fourth part is the last part of the framework. 

A two-dimensional object can be described by a complex refractive index distribution $n\left( {x,y} \right) = 1 - \delta \left( {x,y} \right) + i\beta \left( {x,y} \right)$, where $x$ and $y$ describe the coodinate system of the sample. In differential phase-contrast imaging, one measures the effect of variations of the real part $\delta$ by evaluating the tiny refraction angles of X-rays induced by the specimen with a grating Talbot-Lau interferometer. The differential phase-contrast projection can be expressed by ${\alpha _\theta }\left( s \right) = \partial \left( {\int_l {\delta \left( {x,y} \right)dl} } \right)/\partial s$, where $s$ describes the coordinate system of the detector, $\theta$ the rotation view angle of the object, and $l$ the incident ray direction. Using the three point differential method, the differential phase-contrast projection can be discretely re-expressed into Eq.\eqref{differential}. 

\begin{equation}
\label{differential}
{\alpha _\theta }\left( s \right) \approx \frac{{p\left( {j + 1,m} \right) - p\left( {j - 1,m} \right)}}{2}
\end{equation}

Here, $p$ is a two-dimensional matrix and each element in this matrix represents the line integral value detected by one detector channel. It is used in the third part to get differential phase-contrast sinogram.

Eq.\eqref{fbp} is the well-known two dimensional fan beam FBP algorithm which is adopted to reconstruct the DPC-CT image in this article. It has many extension versions for different CT scanning configurations. In this equation, $\delta \left( {x,y} \right)$ represents the DPC-CT image, $U$ the geometrical weight factor, ${\alpha _\theta }\left( s \right)$ the differential phase-contrast projection sinogram, $h$ the Hilbert filter, Eq.\eqref{hilbert}, $v$ the frequency variant and $\theta$ the rotation angle.

\begin{equation}
\label{fbp}
\delta \left( {x,y} \right) = \frac{1}{2}\int_0^{2\pi } {U \times {\alpha _\theta }} \left( s \right) * h\left( v \right)d\theta
\end{equation}

\begin{equation}
\label{hilbert}
 h\left( v \right) = \frac{1}{{2\pi }}i{\mathop{\rm sgn}} \left( v \right)
\end{equation}

Eq.\eqref{forward_pro} is the forward projection operator. In this equation, $P\left( {\delta \left( {x,y} \right)} \right)$ the forward projection, ${\delta \left( {x,y} \right)}$ represents the initial reconstructed DPC-CT image and $l$ the forward projection path.

\begin{equation}
\label{forward_pro}
P\left( {\delta \left( {x,y} \right)} \right) = \int_0^{2\pi } {\delta \left( {x,y} \right)dl}
\end{equation}

Eq.\eqref{general_neural_network} represents the way how the neural network in this framework produce complete phase-contrast sinogram. Corrupted sinogram $X$ with artifacts is feed into the neural network, then an artifact-free complete phase-contrast sinogram $\hat Y$ could be obtained. Finally, FBP is used again on $\hat Y$ to get the final DPC-CT image.

\begin{equation}
\label{general_neural_network}
\hat Y = F(X)
\end{equation}

\subsection{Neural network}

The neural network is build upon a more elegant architecture, U-net \cite{U-net}, which has wide applications like semantic segmentation \cite{U-net} and medical image processing \cite{convolutionalforCT}, due to its goal is to generate a complete sinogram and the size of input and output are identical. U-net has large number of feature channels to extract high-level features in the extracting path, and large number of feature channels in the expansive path to allow the propagation of context information to higher resolution layers. As a consequence, its symmetric structure is organized in the form of Encoder-Decoder. Recently, research \cite{deepDepth} shows many layers contribute very little and can in fact be randomly dropped during training, and DenseNet \cite{denseNet} points out densely-connection can act as a substitute for large number of feature channels because there is no need to relearn redundant feature-maps. Therefore, this neural network takes fully advantages of U-net and DenseNet to generate complete phase-contrast sinogram. 

According to the spatial resolution of feature maps, this network can be divided into different stages. As depicted in Fig.\ref{framework}, there are five encoding stages which are surrounded by gray dotted boxes, and four decoding stages surrounded by green dotted boxes. Numbers next to each stage denote the corresponding spatial resolution on that stage.

\subsubsection{\bf{Encoder-Decoder}} 

Encoder is used to extract multi scale features from the input corrupted sinogram. Bigger convolutional (Conv.) filter could provide larger receptive field, but more parameters will be introduced, which may cause overfitting and hardly converging. Consecutive convolution with small filter could provide the same size of receptive field as bigger filter does, with less parameters. Therefore, consecutive $3 \times 3$ Conv., Rectified Linear Units (Relu) \cite{relu} and BatchNormalization (BN)\cite{batchnorm} are used to extract features from the input. 

In common use \cite{U-net,ZhichengDLRecon}, max-pooling is taken to reduce the feature maps' spatial resolution to increase the size of receptive field and get multi scale features. It computes fastly, but only keeps the maximum value from the pooling window. In order to pay more attention to important values rather than maximum one, strided convolution is used. The parameters in convolution filter will decide which part is important and give it higher weight. Besides, to reduce the feature maps' spatial resolution as max-pooling does, the stride is set to 2. So $3 \times 3$ Conv. (strides=2) is used for reducing spatial resolution. 

After high-level features are extracted, which possess high semantic information and are abstract, Decoder should make use of those abstract features to restore a high-quality output with the same size as the input corrupted sinogram. Firstly, deconvolution \cite{deconvolution}, which can be regarded as the reverse version of convolution, also known as transpose convolution (ConvTranspose), is adopted to recover the spatial resolution, with stride 2. Then high-level features are concatenated with low-level ones, and $3\times3$ Conv.-Relu-BN are used to refine details. It's working stage-by-stage. When the feature maps' spatial resolution are the same as input's, $1 \times 1$ Conv. is used to merge multi channels into one for matching the input's dimension. 

\subsubsection{\bf{Skip Connection}}
There exists two kinds of skip connections: connection between input and output in the form of residue learning \cite{residuallearning} and connections among different stages. 

Firstly, input is directly added to the output, so the neural network could be reformulated as \eqref{residue_network}. Consequently, it only needs to recognize and remove the artifacts from the corrupted sinogram, rather than removing artifacts and building up the whole complete sinogram from those abstract features at the same time. This way can simplify the learning process, which makes the network focus on the artifact reduction.

\begin{equation}
\label{residue_network}
\hat Y = H\left( X \right) + X
\end{equation}

Secondly, the output from different stages are connected. Though high-level features contains high semantic information, it is not helpful for recovering the spatial information, while low-level features retain spatial accuracy with less semantic information. So low-level features are concatenated with high-level ones to refine the spatial information, which is similar to \cite{U-net,convolutionalforCT,sinoInterpo2,ZhichengDLRecon}.

\subsubsection{\bf{DenseBlock}}

\begin{figure}[!htbp]
\centering

\includegraphics[width=1\columnwidth]{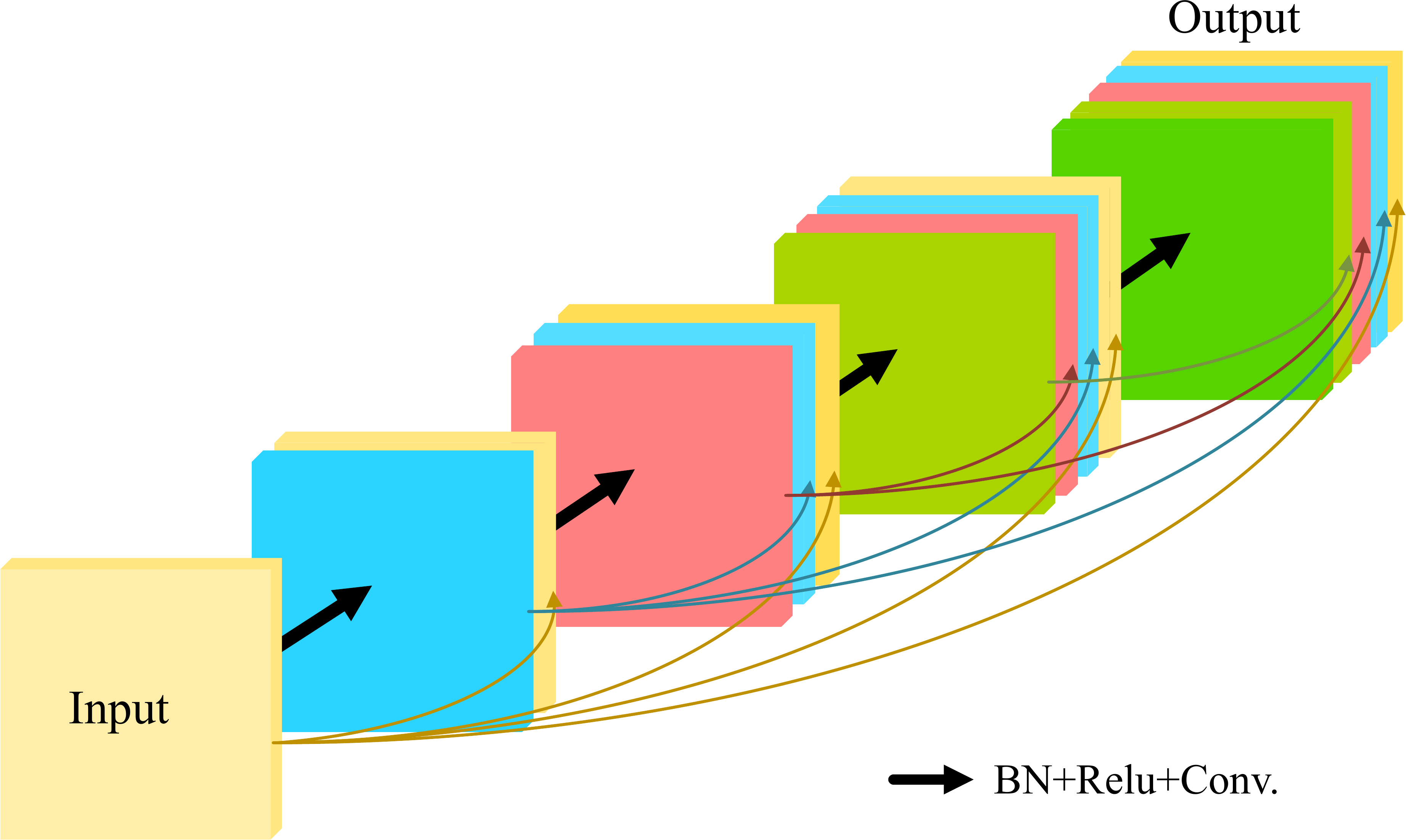}

\caption{The structure of DenseBlock. Arrows with color represent copy and concatenation.}

\label{denseblock}
\end{figure}

DenseBlock is the basic module in DenseNet \cite{denseNet}. It distills skip connection into a simple pattern: all layers are directly connected with each other in the same block. As shown in Fig.\ref{denseblock}, each layer obtains additional inputs from all preceding layers and passes on its own feature maps to all subsequent layers. It exploits the potential of the network through feature reuse instead of extreamly deep or wide architectures, and yield condensed models and highly parameter-efficiency. 

Densely connection makes model easy to train, because each layer in the same block has direct access to the gradients from the loss functions and the original input signal, resulting in implicit deep supervision \cite{deepSupervision}. Further, it proves that small number of feature channels are sufficient to obtain state-of-the-art results, because global information can be reached from everywhere within the network, and unlike in traditional network architectures, there is no need to replicate it from layer to layer. Besides, each layer just adds $k\left( {e.g. \,\,k = 16} \right)$ feature maps to global information, which regulates how much new information each layer contributes to the global one and reduces redundancy. This pattern also has regularizing effect resulting from highly-parameter-efficiency, which reduces overfitting on tasks with smaller training set sizes.

In this paper, each DenseBlock contains four BN-Relu-$5\times5$ Conv. (strides=1) layers, and each layer produces $k$ new feature maps. The parametric architecture of one DenseBlock is shown in Table \ref{dense_parameters}, where $H$ and $W$ denotes spatial resolution of feature maps', ${C_I}$ denotes the number of input feature channels.

\begin{table}[!htbp]
\caption{Parameters in the DenseBlock}
\label{dense_parameters}
\centering
\begin{tabular}{c|c|c}
\hline

\bf{Layers} & \bf{Input Size} & \bf{Output Size} \\ \hline\hline
BN-Relu-$5\times5$ Conv. & $H \!\times \!W \times \!{C_I}$ & $H\! \times\! W\! \times {k}$ \\ \hline
\multirow{2}{*}{Concatenate}   & $H \!\times \!W \times \!{C_I}$    & \multirow{2}{*}{$H \!\times \!W \times \!\left( {{C_I} \!+\! k} \right)$} \\ \cline{2-2}
							   & $H \!\times \!W \times \!{k}$ & \\ \hline
BN-Relu-$5\times5$ Conv. & $H \!\times \!W \times \!\left( {{C_I} \!+\! k} \right)$ & $H \!\times \!W \times \!{k}$ \\ \hline
\multirow{2}{*}{Concatenate}   & $H \!\times \!W \times\! \left( {{C_I} \!+ \!k} \right)$    & \multirow{2}{*}{$H \!\times \!W \times \!\left( {{C_I} \!+ \!2k} \right)$} \\ \cline{2-2}
							   & $H \!\times \!W \times \!{k}$ & \\ \hline
BN-Relu-$5\times5$ Conv. & $H \!\times \!W \times\! \left( {{C_I} \!+ \!2k} \right)$ & $H \!\times \!W \times \!{k}$ \\ \hline
\multirow{2}{*}{Concatenate}   & $H \!\times \!W \times \left( {{C_I} \!+ \!2k} \right)$    & \multirow{2}{*}{$H \!\times \!W \times \!\left( {{C_I}\! + \!3k} \right)$} \\ \cline{2-2}
							   & $H \!\times \!W \times \!{k}$ & \\ \hline
BN-Relu-$5\times5$ Conv. & $H \!\times \!W \times \!\left( {{C_I}\! + \!3k} \right)$ & $H \!\times \!W \times \!{k}$ \\ \hline
\multirow{2}{*}{Concatenate}   & $H \!\times \!W \times \!\left( {{C_I} \!+ \!3k} \right)$    & \multirow{2}{*}{$H \!\times \!W \times\! \left( {{C_I} \!+\! 4k} \right)$} \\ \cline{2-2}
							   & $H \!\times W\! \times \!{k}$ & \\ \hline

\end{tabular} 

\end{table}

\subsubsection{\bf{Parameters in Neural Network}}
The whole architecture of this neural network is shown in Table \ref{nn_parameters}. All the parameters in this neural network are initialized using a Gaussian distribution with zero mean and standard deviation $\sqrt {\frac{2}{{{n_{in}}}}}$ in which ${n_{in}}$ indicates the number of input units in each layer. 

Mean square error (MSE) can force the network to learn difference on pixel value, and Multi-scale structure similarity (MS-SSIM) \cite{ms_ssim} can make it to learn the difference on structure. In this paper, MSE and MS-SSIM are combined to be the loss function, Eq.\eqref{lossfunction}. Here, $\hat Y$ denotes output, and $Y$ represents corresponding grouth truth. The initial learning rate is set to $1 \times {10^{ - 4}}$, and gradually reduced to $1 \times {10^{ - 5}}$.

\begin{equation}
\label{lossfunction}
loss = \frac{1}{2}\left\| {Y - \hat Y} \right\|_2^2 + 1 - MS\_SSIM(Y, \hat Y)
\end{equation}

\begin{table}[!htbp]
\caption{Parameters in the Neural Network}
\label{nn_parameters}
\centering
\begin{tabular}{c|c|c}
\hline

\multirow{2}{*}{\bf{Stages}}   & \multirow{2}{*}{\bf{Layers}}       & \bf{Output Size} \\
								&									& ($k = 16$)  \\ \hline \hline
\multirow{4}{*}{Encoder-1} & $3\times3$ Conv. (strides=1) &    \multirow{2}{*}{$720\times731 \times 2k$} \\
						   & -Relu-BN                     & \\ \cline{2-3}
						   & $3\times3$ Conv. (strides=1) &    \multirow{2}{*}{$720\times731 \times 2k$} \\
						   & -Relu-BN                     & \\ \hline

\multirow{2}{*}{Encoder-2} & $3\times3$ Conv. (strides=2) & $360\times 366 \times 2k$ \\ \cline{2-3}
						   & DenseBlock-1                 & $360 \times 366 \times 6k$ \\ \hline

\multirow{2}{*}{Encoder-3} & $3\times3$ Conv. (strides=2) & $180 \times 183 \times 2k$ \\ \cline{2-3}
						   & DenseBlock-2                 & $180 \times 183 \times 6k$ \\ \hline

\multirow{2}{*}{Encoder-4} & $3\times3$ Conv. (strides=2) & $90 \times 92 \times 2k$ \\ \cline{2-3}
						   & DenseBlock-3                 & $90 \times 92 \times 6k$ \\ \hline

\multirow{2}{*}{Encoder-5} & $3\times3$ Conv. (strides=2) & $45 \times 46 \times 2k$ \\ \cline{2-3}
						   & DenseBlock-4                 & $45 \times 46 \times 6k$ \\ \hline

\multirow{5}{*}{Decoder-1} & $3 \times 3$ ConvTranspose (strides=2) & \multirow{2}{*}{$90 \times 92 \times 6k$} \\
						   & -Relu-BN                     & \\ \cline{2-3}
						   & Concatenate                 & $90 \times 92 \times 12k$ \\ \cline{2-3}
						   & $3\times3$ Conv. (strides=1) &    \multirow{2}{*}{$90 \times 92 \times 6k$} \\ 
						   & -Relu-BN                     & \\ \hline

\multirow{5}{*}{Decoder-2} & $3 \times 3$ ConvTranspose (strides=2) & \multirow{2}{*}{$180 \times 183 \times 6k$} \\
						   & -Relu-BN                     & \\ \cline{2-3}
						   & Concatenate                 & $180 \times 183 \times 12k$ \\ \cline{2-3}
						   & $3\times3$ Conv. (strides=1) &    \multirow{2}{*}{$180 \times 183 \times 6k$} \\ 
						   & -Relu-BN                     & \\ \hline

\multirow{5}{*}{Decoder-3} & $3 \times 3$ ConvTranspose (strides=2) & \multirow{2}{*}{$360 \times 366 \times 6k$} \\
						   & -Relu-BN                     & \\ \cline{2-3}
						   & Concatenate                 & $360 \times 366 \times 12k$ \\ \cline{2-3}
						   & $3\times3$ Conv. (strides=1) &    \multirow{2}{*}{$360 \times 366 \times 6k$} \\ 
						   & -Relu-BN                     & \\ \hline

\multirow{5}{*}{Decoder-4} & $3 \times 3$ ConvTranspose (strides=2) & \multirow{2}{*}{$720 \times 731 \times 6k$} \\
						   & -Relu-BN                     & \\ \cline{2-3}
						   & Concatenate                 & $720 \times 731 \times 8k$ \\ \cline{2-3}
						   & $3\times3$ Conv. (strides=1) &    \multirow{2}{*}{$720 \times 731 \times 4k$} \\ 
						   & -Relu-BN                     & \\ \hline

Merge                      & $1 \times 1$ Conv. (strides=1) & $720 \times 731 \times 1$ \\ \hline
---                        & Add                            & $720 \times 731 \times 1$ \\ \hline

\end{tabular} 

\end{table}

\subsection{Artifact reduction in the sinogram domain}

The output of a neural network can not be entirely correct, and even the latest optimization algorithm \cite{adaboundOp} can not improve the accuracy to 100\%. When DL is applied in the CT image domain, the imprecise value of the output will be directly reflected in the CT image, resulting in the structural deformation. However, Eq. \eqref{fbp} tells that when leveraging FBP on sinogram to obtain CT image, the object information on CT image is obtained by weighted combination of information from all angles on sinogram. Thus, the final CT image reconstructed from the sinogram in which the artifact is reduced by DL could be more tolerant to inaccurate output value, and could get more accurate structure. The experimental section also confirms that the results of artifact reduction conducted in the sinogram domain are better than those whose artifact reduction is conducted on CT image domain.

\subsection{Running modes}
This framework has two running modes. One is training mode and another is working mode. Training mode has following steps: 
\begin{enumerate}
  \item A set of incomplete phase-contrast sinograms is firstly matched with the corresponding complete sinogram into many pairs of training data which include an incomplete sinogram and a complete sinograms. \label{list_1}
  \item These data is fed into the framework depicted in Fig.\ref{framework} one pair by one pair and the network parameters are updated iteratively. \label{list_2}
  \item When all pairs are used once, an outer learning iteration is completed. \label{list_3} 
  \item Repeat steps \ref{list_2} and \ref{list_3} utill the learning converges. \label{list_4}
\end{enumerate}

The procedure for working mode is much simpler. When an incomplete phase-contrast sinogram is fed into the framework determined by training mode, the output of the framework will be a high quality DPC-CT image.

\section{Experiments}

Sparse-view DPC-CT is a typical case with incomplete data. Taking sparse-view DPC-CT as an application example, this section validates the proposed reconstruction framework by synthetic data sets and experimental data sets.

\subsection{Data preparation}

\subsubsection{{\bf{Synthetic}}}
For synthetic data sets, 500 phantoms are used to obtain the fan beam phase-contrast sinograms with different sampling factors. Each phantom consists of tens of ellipses with random refraction coefficients, size, and location. And the sampling factors are set to be 1, 4, 6, 8 and 12. They correspond to 720, 180, 120, 90 and 60 views, respectively. The size of each phantom is $512 \times 512$ pixel. Fan beam sinograms are generated by using the embedded MATLAB function $fanbeam\left( {} \right)$. The width of all the sinograms are 731 pixels. The sinogram with sampling factor 1 has a size $720\times731$ pixel and is treated as complete one. Other sinograms are incomplete. Sinograms of 400 phantoms are used to train the framework and those from another 100 phantoms are used to test the framework.

Within the framework, for each incomplete sinogram, the initial FBP reconstruction is firstly executed with Eq.\eqref{fbp} and Eq.\eqref{hilbert} to obtain the initial DPC-CT image. Then the forward projection operator in Eq.\eqref{forward_pro} and the differential method in Eq.\eqref{differential} are applied to the initial DPC-CT image to generate the corresponding corrupted phase-contrast sinogram with a size $720 \times 731$ pixel. Next the iterative deep learning runs to update the network parameters by making comparison between the corrupted and the complete phase-contrast sinogram.

This framework involves complicated calculations such as image reconstruction, forward projection and convolution. They may lead to unignored computation errors and degrade the training efficiency and accuracy. So we apply normalization operation to input $X$ and ground truth $Y$ to avoid this problem. This normalization operation is expressed in Eq. \eqref{norm} in which $I_n$ represents the normalized image, $I$ the raw image, $mean()$ a operator to obtain mean value and $std()$ a operator to obtain standard deviation value.

\begin{equation}
\label{norm}
I_n = \frac{I-mean(I)}{std(I)}
\end{equation}

\subsubsection{{\bf{Experimental}}}
The experimental CT Lymph Nodes data sets \cite{dataCitation,publicationCitation1,publicationCitation2} is collected from The Cancer Imaging Achive (TCIA) \cite{tciaCitation}. This collection consists of CT images of 90 patients' mediastinum and 86 patients' abdomen. For the consistency of the data sets' size on both synthetic and experimental, 50 out of 86 patients who underwent abdominal imaging are randomly selected, and then split into 40 patients used for training and 10 patients for testing. 10 images are randomly chosen from each of these 50 patients.

\begin{figure}[!htbp]
\centering

\includegraphics[width=1\columnwidth]{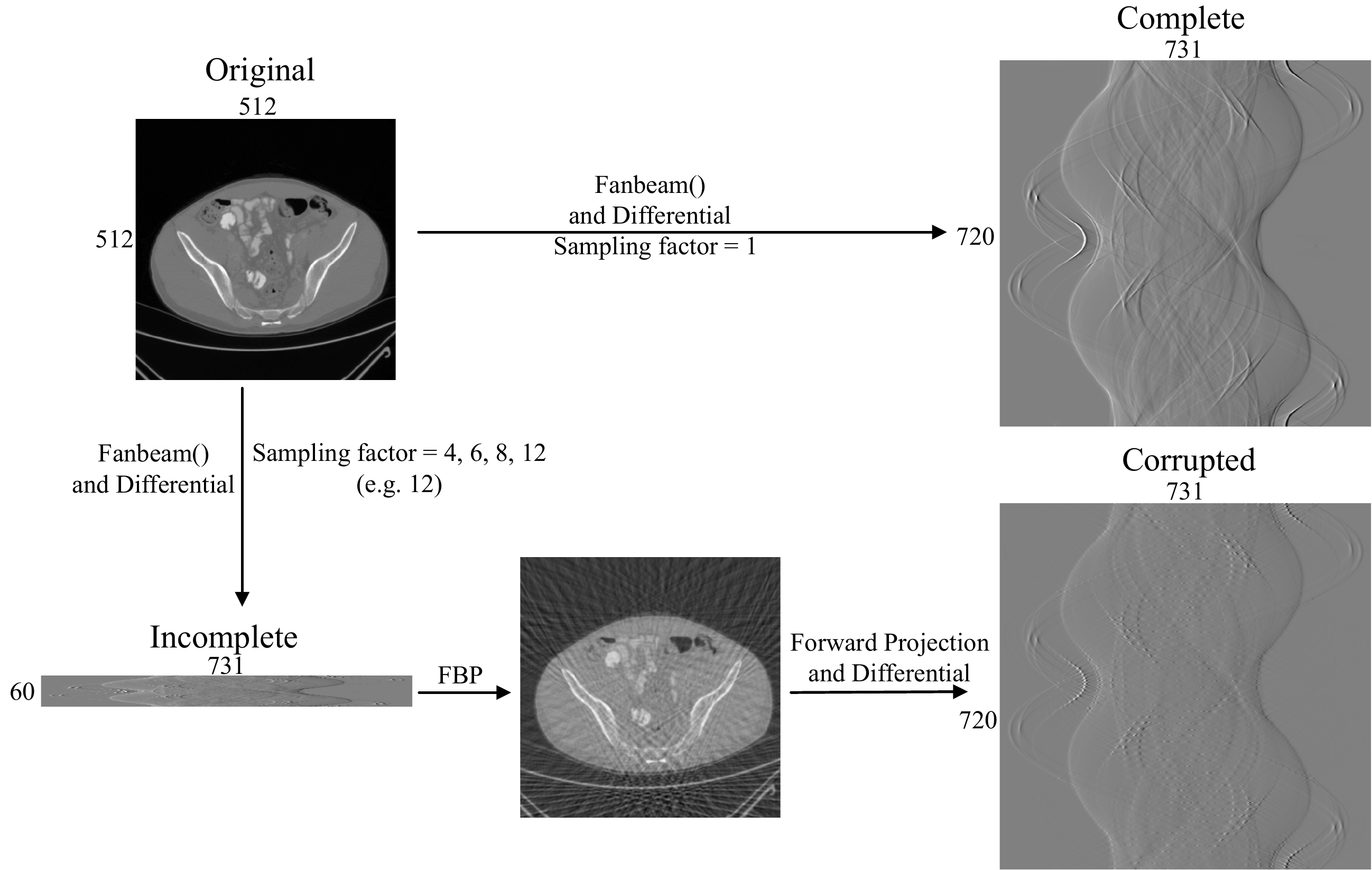}

\caption{An example of how to prepare the training and testing data.}

\label{data_prepare}
\end{figure}

After that, all the operations and procedures are the same as the ones for synthetic data sets. Fig. \ref{data_prepare} is an example of how to prepare the training and testing data.

\subsection{Implementation}
This framework is implemented with Python 3.5.2 and Tensorflow 1.8. It runs in a workstation Advantech AIMB-785 with a CPU i7 6700 and a Graphics Processing Unit (GPU) nVidia GTX 1080Ti 11 GBytes.

For synthetic data sets, datas from 60, 90, 120 and 180 views are used to train a model seperately. So experimental data sets do. 

Adam \cite{adam} algorithm is used to train the neural network. The mini-batch size is 2. All the models are trained for 100 epochs.

\subsection{Comparison with other methods}

In this study, two methods (deep-neural-network-enabled sinogram synthesis, DNN-SS \cite{sinoInterpo2} and sinogram-normalization interpolation, SN-I \cite{analyticInterpo}) based on sinogram completion and one method (DenseNet-Deconvolution network, DD-Net \cite{ZhichengDLRecon}) based on post-processing are conducted to compare the performance with the proposed framework.

\subsection{Image evaluation}

When training, MSE and MS-SSIM have been used as the loss function to guide the network to optimize parameters. Consequently, the network will try to generate output that has low MSE and high MS-SSIM index. So MSE and MS-SSIM are not objective and not suitable for quantitative measurement of the network performance. Besides, Lin Zhang et al \cite{fullReference} compared the performance of several full reference image evaluation methods, such as feature similarity (FSIM) \cite{fsim}, MS-SSIM, SSIM \cite{ssim} and peak signal to noise ratio (PSNR), and results show that FSIM and information content weighted SSIM index (IW-SSIM) \cite{iw_ssim} are more accurate than others. Therefore, FSIM and IW-SSIM are used as quantitative metrics.

Qualitative evaluation is also carried out using visual inspections and image intensity profiles. 

Relative improvement ratio based on FSIM and IW-SSIM index using different methods compared to FBP is calculated with Eq.\eqref{relativeIndex}.

\begin{equation}
\label{relativeIndex}
relI = \frac{{Met - MetonFBP}}{{MetonFBP}}
\end{equation}

Where $relI$ represent relative improvement ratio, $Met$ denotes metric value using FSIM and IW-SSIM on different methods. $MetonFBP$ denotes metric value using FSIM and IW-SSIM on FBP images.

\subsection{Results}

\subsubsection{Synthetic data sets}

\begin{figure}[!htbp]
\centering
\includegraphics[width=1\columnwidth]{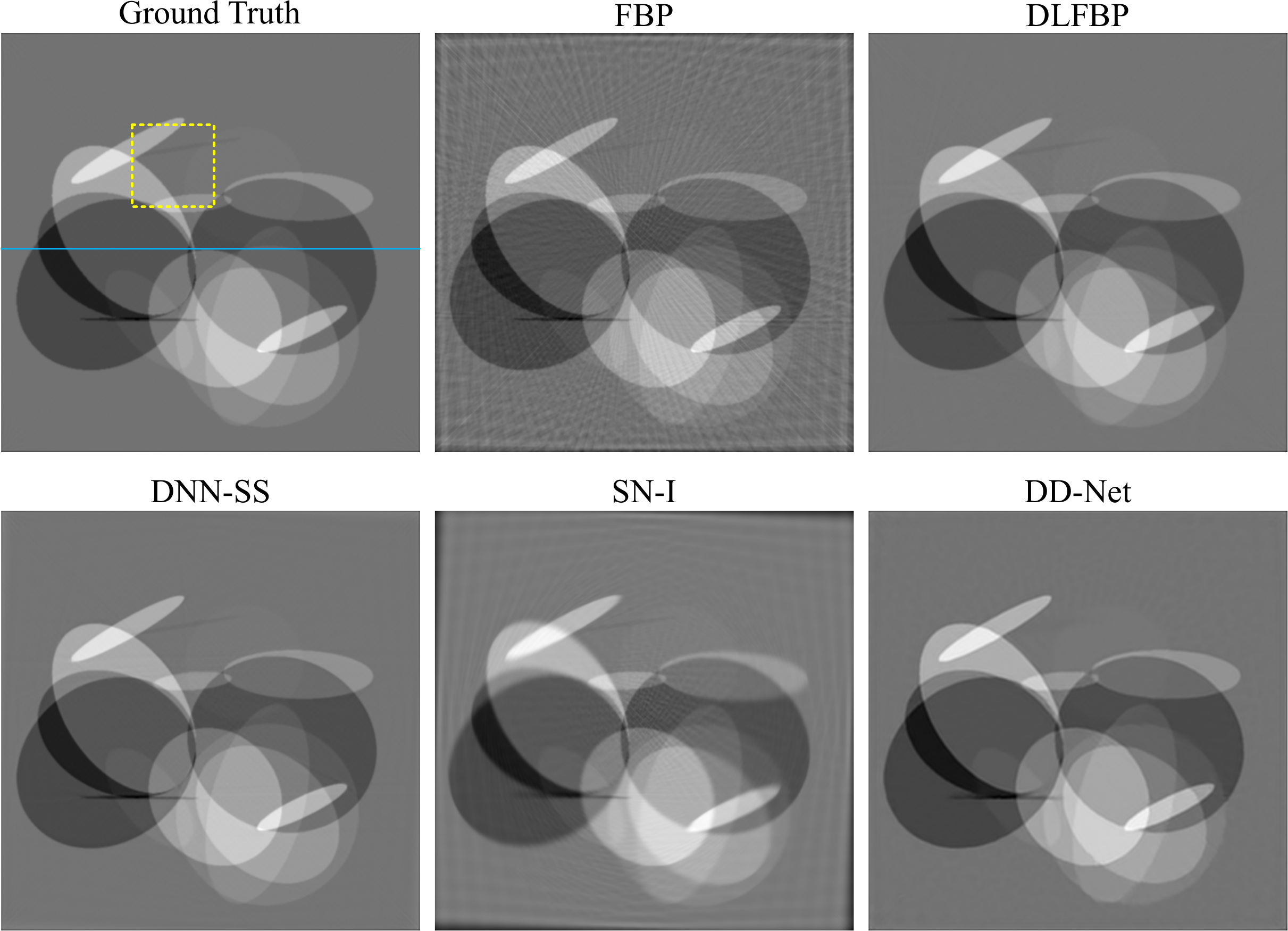}
\caption{One group of results from synthetic data sets with incomplete phase-contrast sinogram of 120 views, using FBP, DLFBP, DNN-SS, SN-I and DD-Net. Same regions of these images, indicated by yellow box, are enlarged for better visualization.}
\label{syn_p_120}
\end{figure}

\begin{figure}[!htbp]
\centering
\includegraphics[width=1\columnwidth]{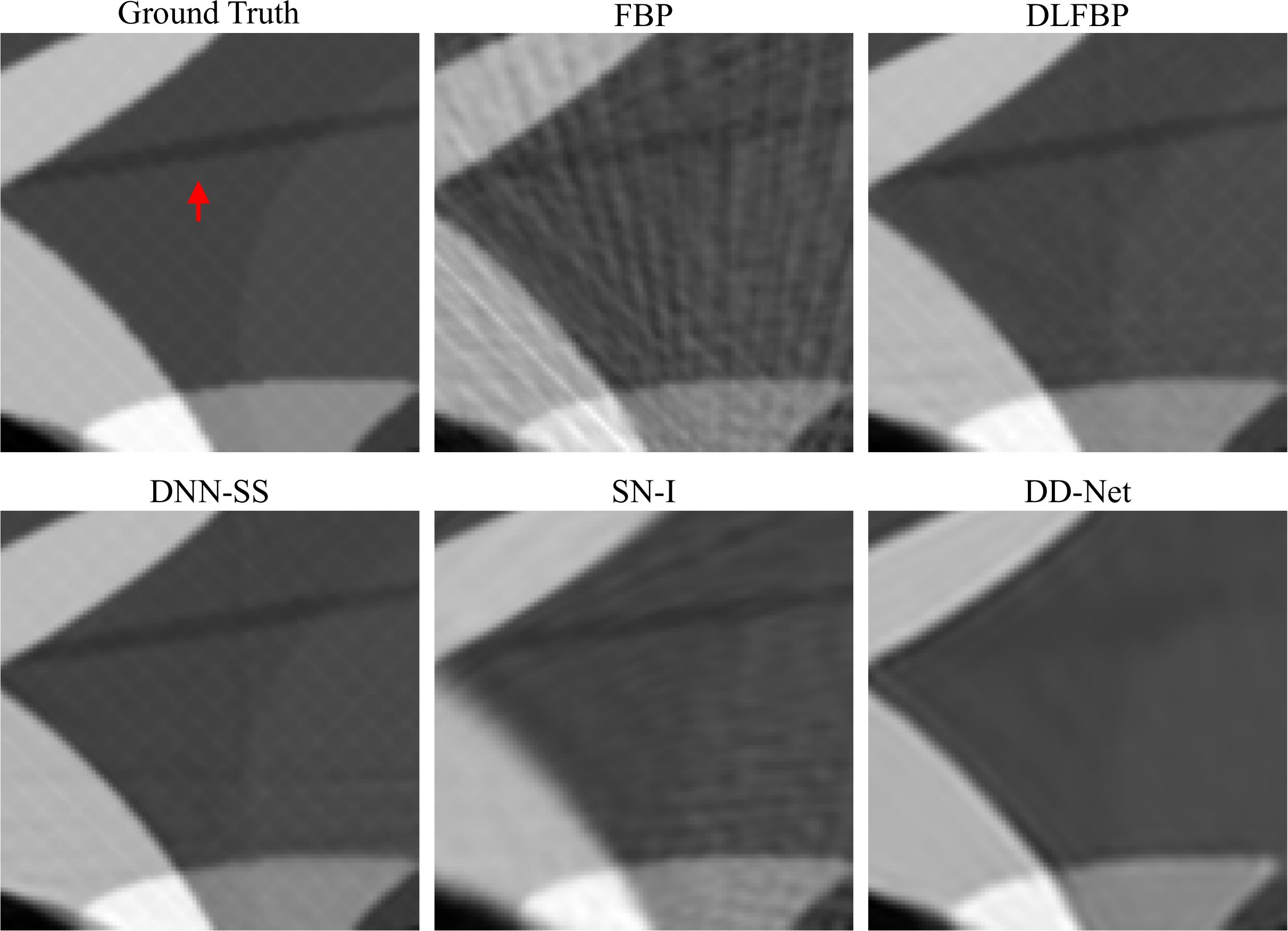}
\caption{Enlarged regions indicated by yellow box in Fig.\ref{syn_p_120} for better visualization. DLFBP and DNN-SS could remove the artifacts and retain small image structures.}
\label{syn_p_120_roi}
\end{figure}

\begin{figure}[!htbp]
\centering
\includegraphics[width=1\columnwidth]{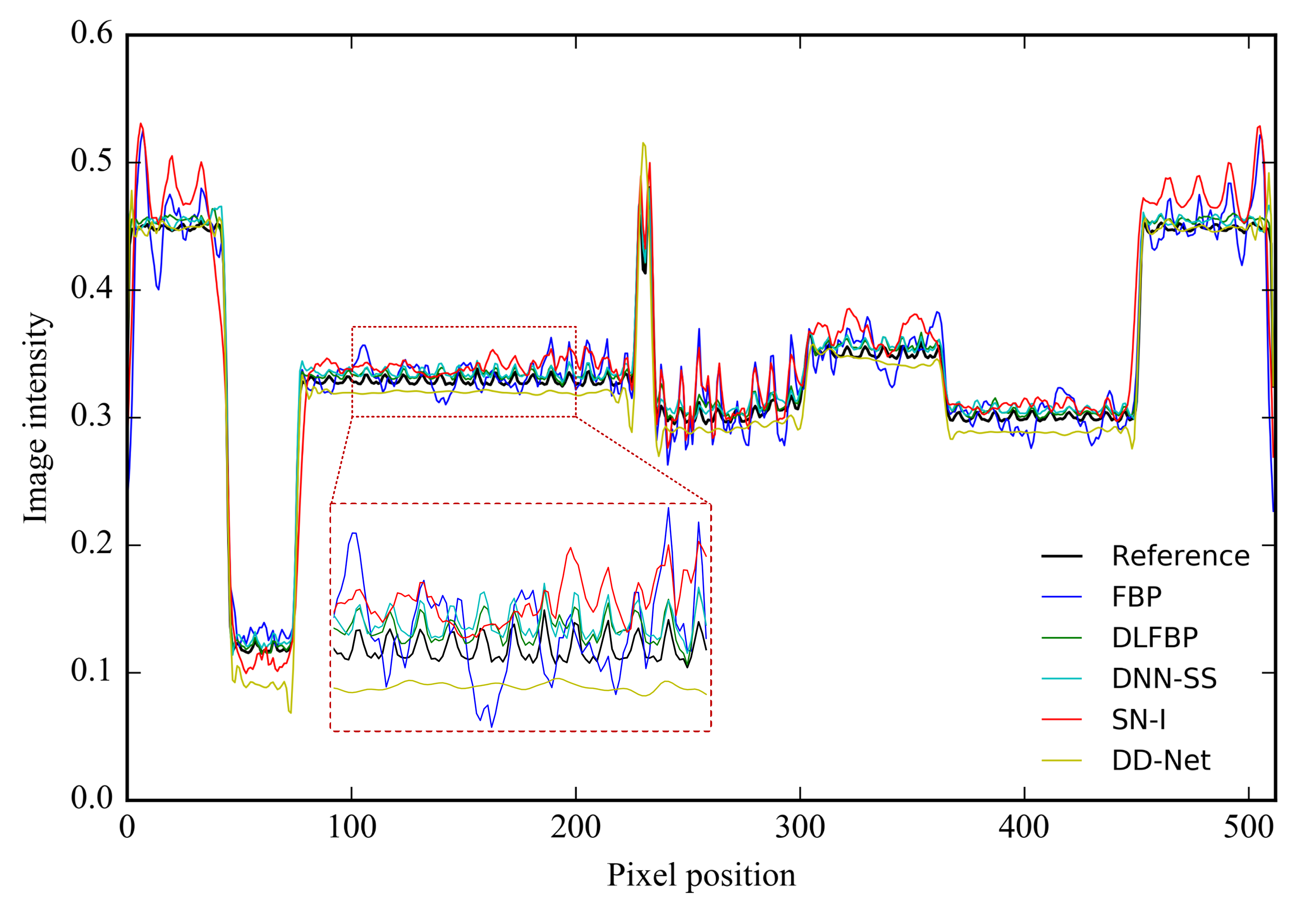}
\caption{Image intensity profiles as measured along the blue solid line in Fig.\ref{syn_p_120}. The profile using DLFBP is the closest to that of the reference image.}
\label{syn_line}
\end{figure}

Figs.\ref{syn_p_120}-\ref{syn_line} present results of one of the 100 synthetic phantoms with 120 views for testing using different methods. Same regions indicated with yellow box are enlarged for better visualization in Fig.\ref{syn_p_120_roi}. The image intensity profile in same position is shown in Fig.\ref{syn_line}. 

As expected, severe artifacts exist in the result using FBP, and DLFBP, DNN-SS, SN-I and DD-Net could reduce the artifacts. Though most artifacts are suppressed, there still remains some in the background using SN-I, furthermore, its central part of the image has the clearest structure and the nearer the boundary, the more blurred it is. DD-Net could also remove the artifacts satisfactorily, but the result is overly smoothed, and there is slight distortion in the edge structure. As shown in Fig.\ref{syn_p_120_roi} and Fig.\ref{syn_line}, DD-Net loses small image details and the pixel value changes gently and far from the reference value. Through visual inspection, DLFBP and DD-Net could remove the artifacts and retain small structures. 

Table \ref{tab_theSyn} shows the quantitative measurement using FSIM and IW-SSIM index on this synthetic phantom. DLFBP gets higher value on FSIM and IW-SSIM index compared to other methods.

\begin{figure}[!htbp]
\centering
\includegraphics[width=0.9\columnwidth]{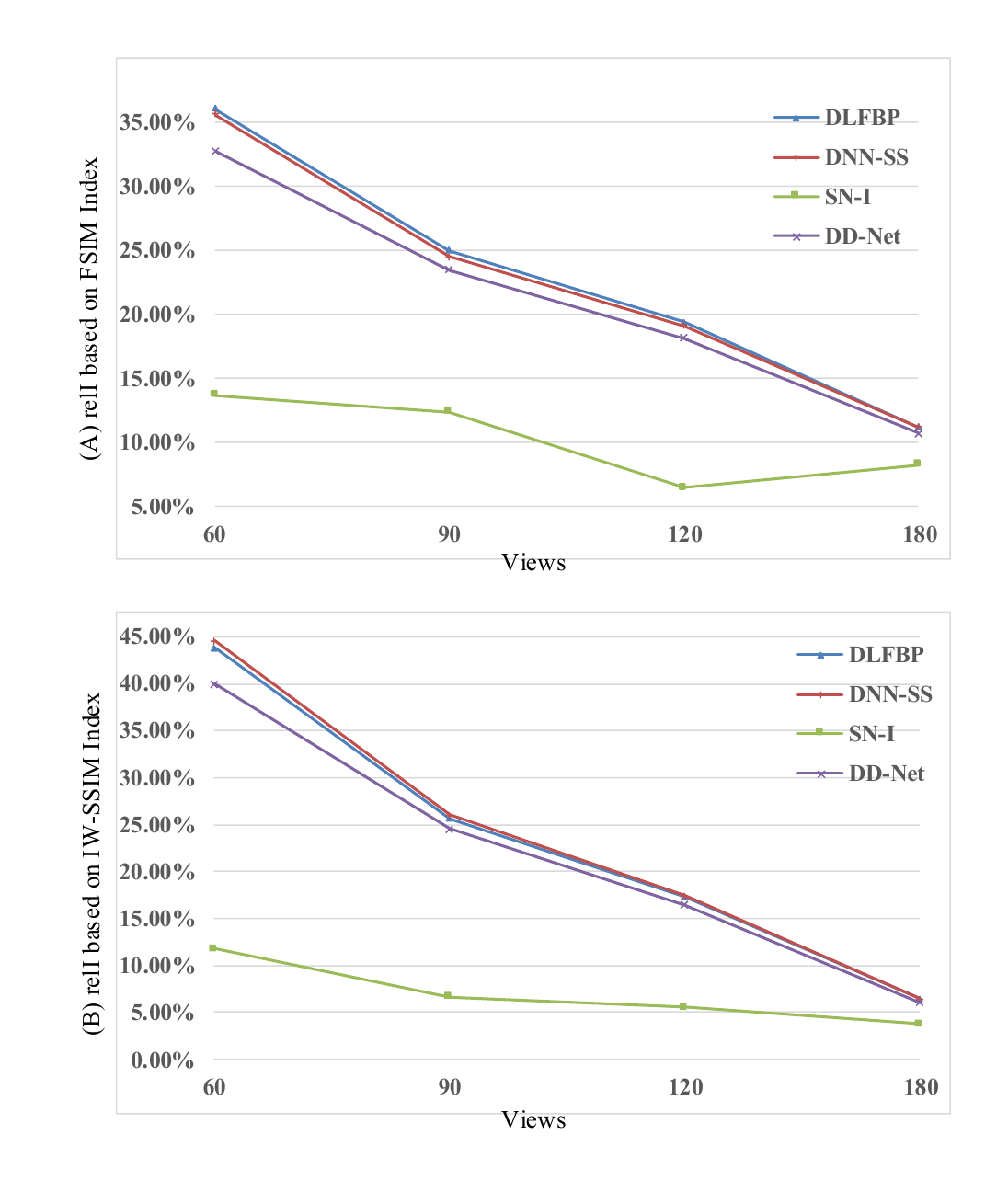}
\caption{Relative improvement ratio based on FSIM and IW-SSIM index using different methods compared to FBP on synthetic data sets. (A) Relative improvement ratio based on FSIM index. (B) Relative improvement ratio based on IW-SSIM index.}
\label{syn_metric}
\end{figure}

Fig.\ref{syn_metric} shows the average relative improvement ratio based on FSIM and IW-SSIM index using different methods compared to the results from FBP on the whole 100 synthetic testing datas. Tabel \ref{tab_syn_p} lists the average FSIM and IW-SSIM values using different methods on 60, 90, 120, and 180 views' synthetic data sets. DLFBP and DNN-SS get higher metric values than other methods.

\begin{table}[!htbp]
\caption{Quantitative measurement using different methods for this synthetic phantom.}
\label{tab_theSyn}
\centering

\begin{tabular}{c|c|c|c|c|c}
\hline
     & FBP & DLFBP & DNN-SS & SN-I & DD-Net \\ \hline \hline
FSIM & 0.841	& \bf{0.995} & 0.994 & 0.872 & 0.986 \\ \hline
IW-SSIM & 0.866 & \bf{0.996} & 0.996 & 0.892 & 0.986 \\ \hline

\end{tabular}
\end{table}

\begin{table}[!htbp]
\caption{Quantitative measurement using different methods for synthetic data sets.}
\label{tab_syn_p}
\centering

\begin{tabular}{c|c|c|c|c|c}
\hline
                      & Methods & 60 & 90 & 120 & 180 \\ \hline\hline
\multirow{5}{*}{FSIM} & FBP     & 0.718 & 0.790 & 0.831 & 0.897 \\ \cline{2-6}
					  & DLFBP   & \bf{0.977} & \bf{0.988} & \bf{0.992} & \bf{0.998} \\ \cline{2-6}
					  & DNN-SS  & 0.974 & 0.984 & 0.990 & 0.997 \\ \cline{2-6}
					  & SN-I    & 0.816 & 0.888 & 0.885 & 0.971 \\ \cline{2-6}
					  & DD-Net  & 0.953 & 0.976 & 0.982 & 0.993 \\ \hline
\multirow{5}{*}{IW-SSIM} & FBP  & 0.678 & 0.788 & 0.848 & 0.938 \\ \cline{2-6}
						 & DLFBP & 0.975 & 0.990 & 0.995 & \bf{0.998} \\ \cline{2-6}
						 & DNN-SS & \bf{0.980} & \bf{0.993} & \bf{0.996} & 0.998 \\ \cline{2-6}
						 & SN-I & 0.758 & 0.840 & 0.895 & 0.974 \\ \cline{2-6}
						 & DD-Net & 0.949 & 0.981 & 0.988 & 0.995 \\ \hline
\end{tabular}
\end{table}

\subsubsection{Experimental data sets}
\begin{figure}[!htbp]
\centering
\includegraphics[width=1\columnwidth]{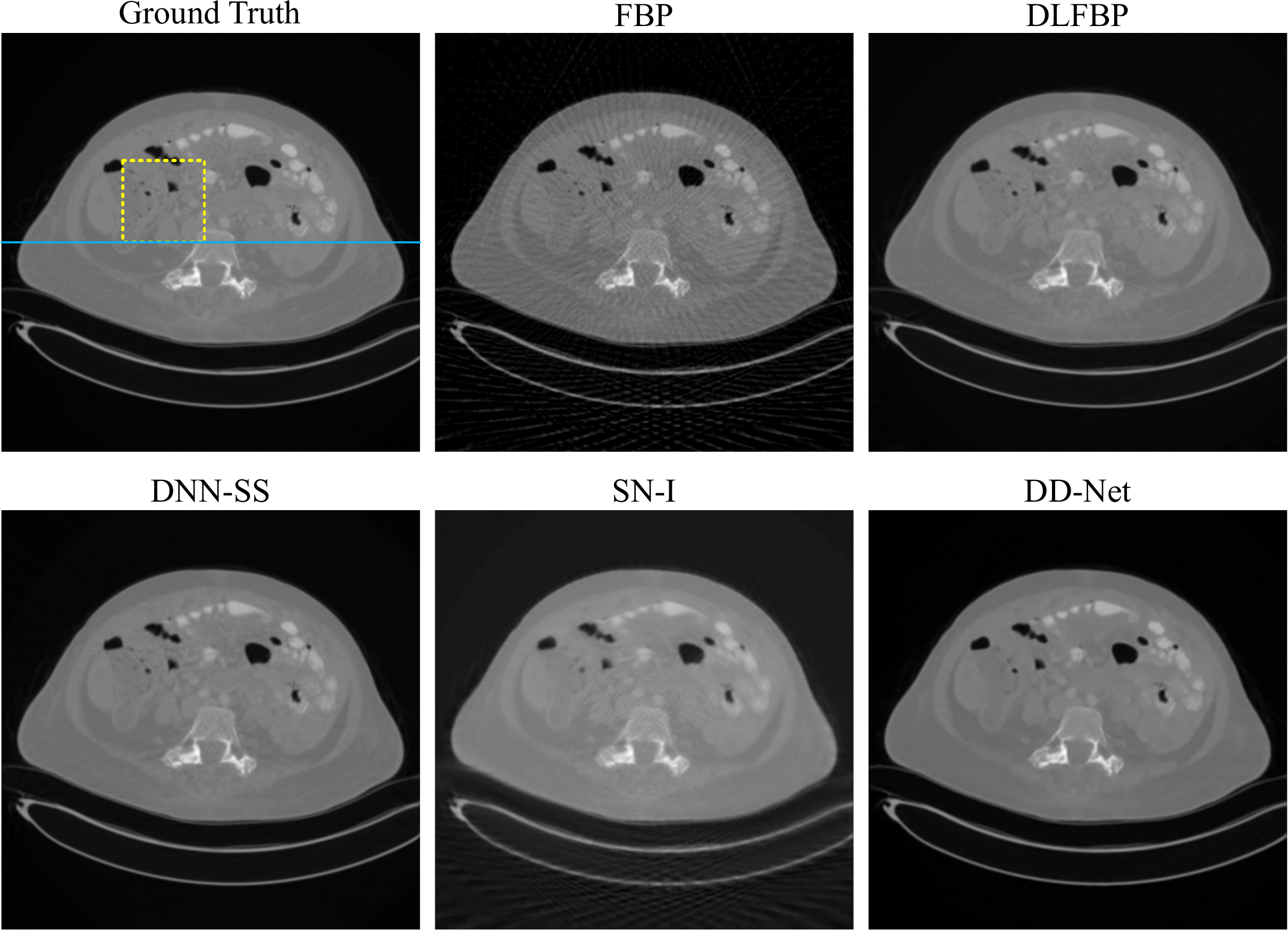}
\caption{One group of results from experimental data sets with incomplete phase-contrast sinogram of 120 views, using FBP, DLFBP, DNN-SS, SN-I and DD-Net. Same regions of these images, indicated by yellow box, are enlarged for better visualization.}
\label{ex_p_120}
\end{figure}

\begin{figure}[!htbp]
\centering
\includegraphics[width=1\columnwidth]{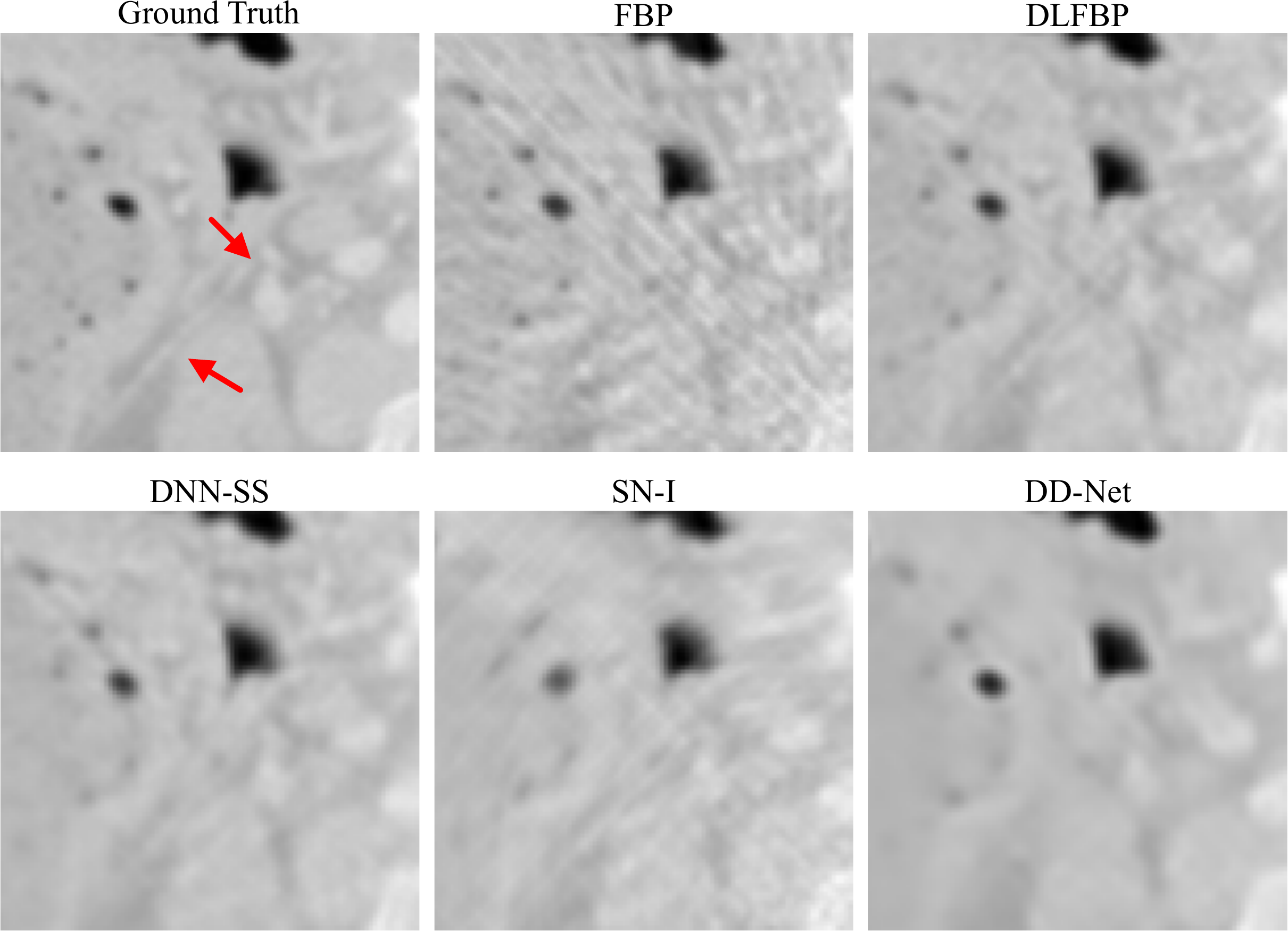}
\caption{Enlarged regions indicated by yellow box in Fig.\ref{ex_p_120} for better visualization. DLFBP and DNN-SS could remove the artifacts and keep tiny structures.}
\label{ex_p_120_roi}
\end{figure}

\begin{figure}[!htbp]
\centering
\includegraphics[width=1\columnwidth]{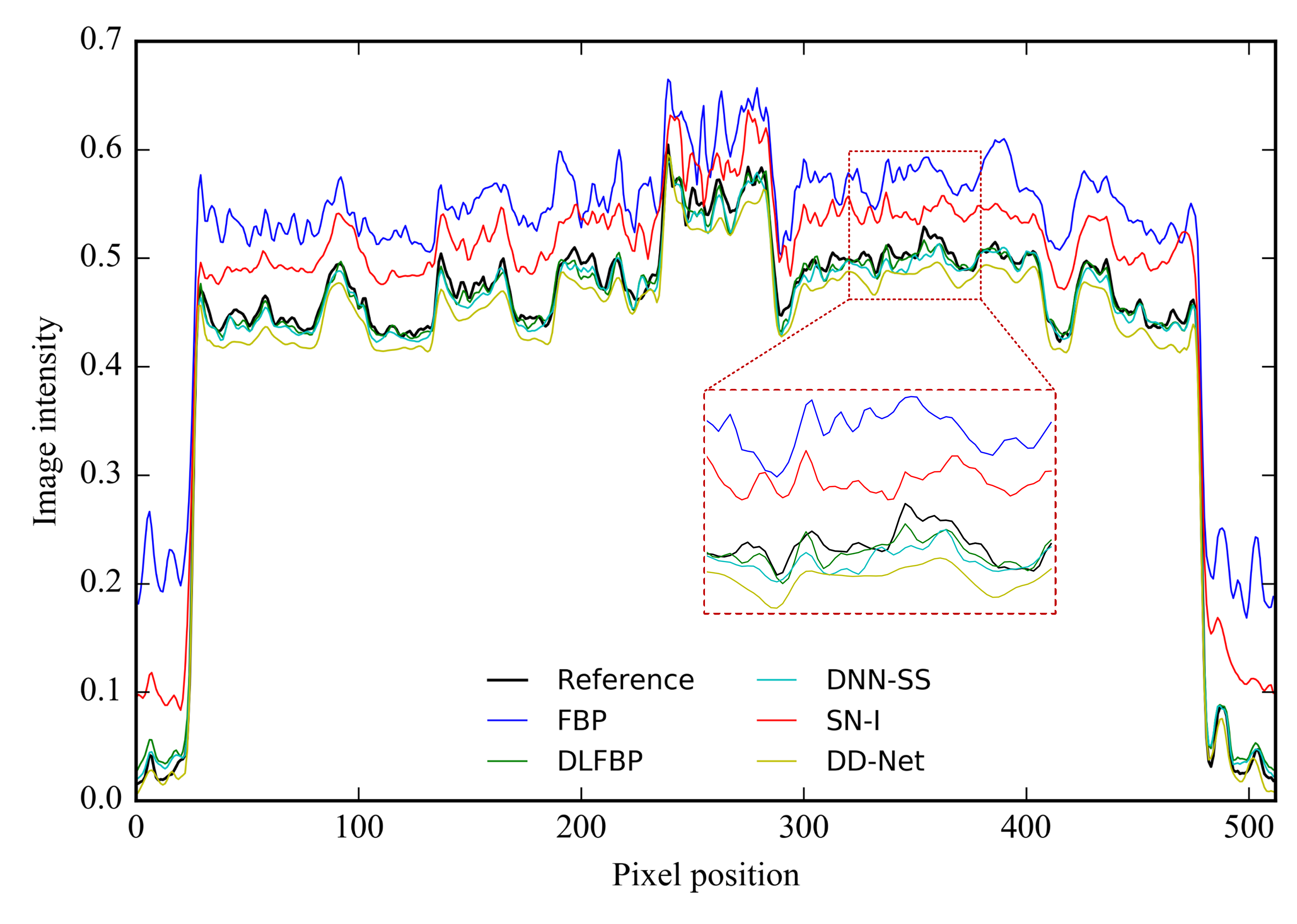}
\caption{Image intensity profiles as measured along the blue solid line in Fig.\ref{ex_p_120}. The profile using DLFBP is the closest to that of the reference image.}
\label{exp_line}
\end{figure}

Figs.\ref{ex_p_120}-Fig.\ref{exp_line} present results of one of the 100 experimental abdominal slice with 120 views for testing using different methods. Same regions indicated with yellow box are enlarged for better visualization in Fig.\ref{ex_p_120_roi}. The image intensity profile in same position is shown in Fig.\ref{exp_line}. 

Still, SN-I could not completely remove the artifacts, and its edge is blurred. As shown in Fig.\ref{ex_p_120_roi} and Fig.\ref{exp_line}, DLFBP, DNN-SS and DD-Net could suppress the artifacts clearly, while the result from DD-Net is overly smoothed and loses details.

Table \ref{tab_theExp} shows the quantitative measurement using FSIM and IW-SSIM index on this experimental slice. DLFBP and DNN-SS get higher value on FSIM and IW-SSIM index compared to other methods.

\begin{figure}[!htbp]
\centering
\includegraphics[width=0.9\columnwidth]{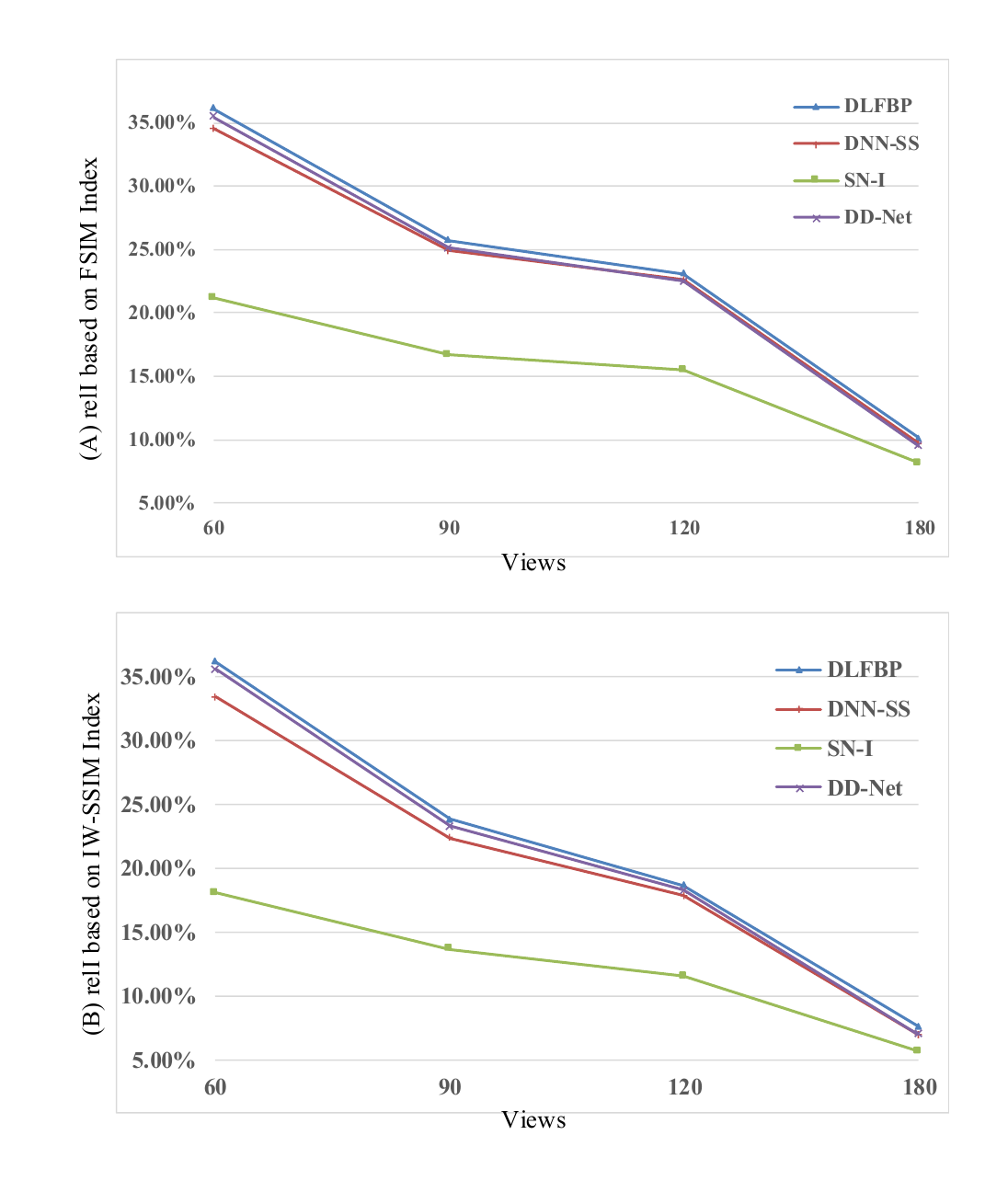}
\caption{Relative improvement ratio based on FSIM and IW-SSIM index using different methods compared to FBP on experimental data sets. (A) Relative improvement ratio based on FSIM index. (B) Relative improvement ratio based on IW-SSIM index.}
\label{exp_metric}
\end{figure}

Fig.\ref{exp_metric} shows the average relative improvement ratio based on FSIM and IW-SSIM index using different methods compared to the results from FBP on the entire experimental testing data sets. Tabel \ref{tab_ex_p} lists the average FSIM and IW-SSIM values using different methods on 60, 90, 120, and 180 views' experimental testing data sets. DLFBP and DNN-SS get higher metric values than other methods.

\begin{table}[!htbp]
\caption{Quantitative measurement using different methods for the experimental data.}
\label{tab_theExp}
\centering
\begin{tabular}{c|c|c|c|c|c}
\hline
         & FBP & DLFBP & DNN-SS & SN-I & DD-Net \\ \hline \hline
FSIM    & 0.801	& \bf{0.990} & 0.990 & 0.930 & 0.987 \\ \hline
IW-SSIM & 0.835 & \bf{0.990} & 0.989 & 0.931 & 0.987 \\ \hline
\end{tabular}
\end{table}

\begin{table}[!htbp]
\caption{Quantitative measurement using different methods for experimental data sets.}
\label{tab_ex_p}
\centering

\begin{tabular}{c|c|c|c|c|c}
\hline
                      & Methods & 60 & 90 & 120 & 180 \\ \hline\hline
\multirow{5}{*}{FSIM} & FBP     & 0.712 & 0.782 & 0.802 & 0.904 \\ \cline{2-6}
					  & DLFBP   & \bf{0.969} & \bf{0.983} & \bf{0.988} & \bf{0.996} \\ \cline{2-6}
					  & DNN-SS  & 0.958 & 0.976 & 0.984 & 0.992 \\ \cline{2-6}
					  & SN-I    & 0.862 & 0.912 & 0.927 & 0.978 \\ \cline{2-6}
					  & DD-Net  & 0.965 & 0.978 & 0.983 & 0.989 \\ \hline
\multirow{5}{*}{IW-SSIM} & FBP  & 0.705 & 0.793 & 0.832 & 0.926 \\ \cline{2-6}
						 & DLFBP & \bf{0.961} & \bf{0.982} & \bf{0.988} & \bf{0.996} \\ \cline{2-6}
						 & DNN-SS & 0.941 & 0.970 & 0.981 & 0.990 \\ \cline{2-6}
						 & SN-I & 0.833 & 0.901 & 0.928 & 0.978 \\ \cline{2-6}
						 & DD-Net & 0.957 & 0.978 & 0.985 & 0.990 \\ \hline
\end{tabular}
\end{table}

\subsubsection{Parameters and computational cost}

The synthetic and experimental results show DLFBP and DNN-SS outperform other methods in artifacts reduction and structure preservation. Using FSIM and IW-SSIM index as quantitative measurement, these two get high value than others. Besides, DLFBP and DNN-SS both corrected the missing information in the phase-contrast sinogram domain. But the number of parameters used in DNN-SS is 27 times that of DLFBP. And when inference with a single sinogram, DNN-SS will spend more time than DLFBP. Table \ref{tab_para} shows the number of parameters in DNN-SS and DLFBP, and the inference time using a single sinogram.

\begin{table}[!htbp]
\caption{Parameters and computational cost in DNN-SS and DLFBP.}
\label{tab_para}
\centering
\begin{tabular}{c|c|c}
\hline
         & DNN-SS & DLFBP \\ \hline \hline
Number of parameters (million) & 37.50 & \bf{1.36} \\ \hline
Inference Time (seconds per sinogram) & 3.56 & \bf{0.21} \\ \hline
\end{tabular}
\end{table}

\section{Conclusion}

In this paper, we reported a new deep learning reconstruction framework for DPC-CT with incomplete data. Different from post-processing and sinogram completion methods, the artifacts are firstly projected into corrupted sinogram, in which the missing information will be completed in a fasion of CT scanning instead of numerical interpolation, and then it is reduced by a neural network. Validated with synthetic and experimental data sets, it could remove artifacts clearly and preserve structures at the same time. 

Compared with sinogram completion and post-processing methods and taking FSIM and IW-SSIM index as quantitative metrics, the results from which missing information are corrected in the phase-contrast sinogram domain are better than those from post-processing method. Post-processing will lose tiny details and deform structures, which may be caused by the imprecise output from neural network. When artifacts reduction is conducted in the sinogram domain, it is more tolerant to the inaccurate output from the neural network, because the objects are weighted combination of all projections in the sinogram.

In addition, this framework has high computational efficiency. The neural network is constructed based on elegant models, U-net and DenseNet. U-net could extract multi-scale features from the input and then make use of those abstract features to generate complete sinogram. But there are large number of feature channels in original U-net, and DenseNet has demonstrated small number of feature channels is sufficient to achieve the same effect through densely connections. By combination of U-net and DenseNet, the reported framework could get slightly better results while the number of parameters is 27 times less than that of the method which also corrected information in the sinogram domain.

\section*{Acknowledgment}
We acknowledge support from National Natural Science Foundation of China (12014003), the large scale science facilities joint funed by National Natural Science Foundation of China and Chinese Academy of Science (U1432101, 11179009).

\bibliography{IEEEabrv,thereference}

\begin{thebibliography}{10}
\providecommand{\url}[1]{#1}
\csname url@samestyle\endcsname
\providecommand{\newblock}{\relax}
\providecommand{\bibinfo}[2]{#2}
\providecommand{\BIBentrySTDinterwordspacing}{\spaceskip=0pt\relax}
\providecommand{\BIBentryALTinterwordstretchfactor}{4}
\providecommand{\BIBentryALTinterwordspacing}{\spaceskip=\fontdimen2\font plus
\BIBentryALTinterwordstretchfactor\fontdimen3\font minus
  \fontdimen4\font\relax}
\providecommand{\BIBforeignlanguage}[2]{{%
\expandafter\ifx\csname l@#1\endcsname\relax
\typeout{** WARNING: IEEEtran.bst: No hyphenation pattern has been}%
\typeout{** loaded for the language `#1'. Using the pattern for}%
\typeout{** the default language instead.}%
\else
\language=\csname l@#1\endcsname
\fi
#2}}
\providecommand{\BIBdecl}{\relax}
\BIBdecl

\bibitem{phase1}
R.~Fitzgerald, ``Phase‐sensitive x‐ray imaging,'' \emph{Physics Today},
  vol.~53, no.~7, pp. 23--26, 2000.

\bibitem{phase2}
A.~Momose, ``Phase-sensitive imaging and phase tomography using x-ray
  interferometers,'' \emph{Optics Express}, vol.~11, no.~19, pp. 2303--2314,
  2003.

\bibitem{phase3}
U.~Bonse and M.~Hart, ``An x‐ray interferometer,'' \emph{Applied Physics
  Letters}, vol.~6, no.~8, pp. 155--156, 1965.

\bibitem{phase4}
A.~Momose, T.~Takeda, Y.~Itai, and K.~Hirano, ``Phase–contrast x–ray
  computed tomography for observing biological soft tissues,'' \emph{Nature
  Medicine}, vol.~2, pp. 473--475, 1996.

\bibitem{phase5}
V.~N. Ingal and E.~A. Beliaevskaya, ``X-ray plane-wave topography observation
  of the phase contrast from a non-crystalline object,'' \emph{Journal of
  Physics D: Applied Physics}, vol.~28, no.~11, p. 2314, 1995.

\bibitem{phase6}
T.~Davis, D.~Gao, T.~Gureyev, A.~Stevenson, and S.~Wilkins, ``Phase-contrast
  imaging of weakly absorbing materials using hard x-rays,'' \emph{Nature},
  vol. 373, pp. 595--598, 1995.

\bibitem{phase7}
D.~Chapman, W.~Thomlinson, R.~Johnston, and D.~Sayers, ``Diffraction enhanced
  x-ray imaging,'' \emph{Physics in Medicine and Biology}, vol.~42, no.~11, pp.
  2015--2025, 1997.

\bibitem{phase8}
A.~Snigirev, I.~Snigireva, V.~Kohn, and S.~Igor, ``On the possibilities of
  x‐ray phase contrast microimaging by coherent high‐energy synchrotron
  radiation,'' \emph{Review of Scientific Instruments}, vol.~66, no.~12, pp.
  5486--5492, 1995.

\bibitem{phase9}
S.~Wilkins, T.~Gureyev, D.~Gao, and A.~W. Stevenson, ``Phase-contrast imaging
  using polychromatic hard x-rays,'' \emph{Nature}, vol. 384, no. 6607, pp.
  335--338, 1996.

\bibitem{phase10}
W.~Ludwig, J.~Barichel, D.~van dyck, and J.-P. Guigay, ``Holotomography:
  Quantitative phase tomography with micrometer resolution using hard
  synchrotron radiation x rays,'' \emph{Applied Physics Letters}, vol.~75,
  no.~19, 1999.

\bibitem{phase11}
K.~A. Nugent, T.~Gureyev, D.~Cookson, and Z.~Barnea, ``Quantitative phase
  imaging using hard x rays,'' \emph{Physical Review Letters}, vol.~77, no.~14,
  pp. 2961--2964, 1996.

\bibitem{phase12}
A.~Momose, S.~Kawamoto, I.~Koyama, and Y.~Suzuki, ``Demonstration of x-ray
  talbot interferometry,'' \emph{Japanese Journal of Applied Physics}, vol.~42,
  no.~2, pp. L866--L868, 2003.

\bibitem{phase13}
T.~Weitkamp, A.~Diaz, C.~David, F.~Pfeiffer \emph{et~al.}, ``X-ray phase
  imaging with a grating interferometer,'' \emph{Optics Express}, vol.~13,
  no.~16, pp. 6296--6304, 2005.

\bibitem{phase14}
A.~Momose, ``Recent advances in x-ray phase imaging,'' \emph{Japanese Journal
  of Applied Physics}, vol.~44, no.~9A, p. 6355, 2005.

\bibitem{phase15}
F.~Pfeiffer, T.~Weitkamp, O.~Bunk, and C.~David, ``Phase retrieval and
  differential phase-contrast imaging with low-brilliance x-ray sources,''
  \emph{Nature Physics}, vol.~2, no.~4, pp. 258--261, 2006.

\bibitem{phase16}
F.~Pfeiffer, C.~Kottler, O.~Bunk, and C.~David, ``Hard x-ray phase tomography
  with low-brilliance sources,'' \emph{Physical Review Letters}, vol.~98,
  no.~10, 2007.

\bibitem{phase17}
F.~Pfeiffer, M.~Bech, O.~Bunk \emph{et~al.}, ``Hard-x-ray dark-field imaging
  using a grating interferometer,'' \emph{Nature Materials}, vol.~7, no.~2, pp.
  134--137, 2008.

\bibitem{phase18}
M.~Bech, T.~H. Jensen, R.~Feidenhansl, O.~Bunk \emph{et~al.}, ``Soft-tissue
  phase-contrast tomography with an x-ray tube source,'' \emph{Physics in
  Medicine and Biology}, vol.~54, no.~9, pp. 2747--2753, 2009.

\bibitem{phase19}
T.~Donath, F.~Pfeiffer, O.~Bunk \emph{et~al.}, ``Toward clinical x-ray
  phase-contrast ct: demonstration of enhanced soft-tissue contrast in human
  specimen,'' \emph{Investigative Radiology}, vol.~45, no.~7, pp. 445--452,
  2010.

\bibitem{phase21}
Z.-F. Huang, K.-J. Kang, L.~Zhang, Z.-Q. Chen \emph{et~al.}, ``Alternative
  method for differential phase-contrast imaging with weakly coherent hard x
  rays,'' \emph{Physical Review A}, vol.~79, no.~1, 2009.

\bibitem{phase20}
A.~Olivo and R.~D. Speller, ``A coded-aperture technique allowing x-ray phase
  contrast imaging with conventional sources,'' \emph{Applied Physics Letters},
  vol.~91, no.~7, 2007.

\bibitem{phase22}
Z.~Wang, P.~Zhu, W.~Huang, Q.~Yuan, X.~Liu, K.~Zhang \emph{et~al.}, ``Analysis
  of polychromaticity effects in x-ray talbot interferometer,''
  \emph{Analytical and Bioanalytical Chemistry}, vol. 397, no.~6, pp.
  2137--2141, 2010.

\bibitem{phase23}
Z.~Wang, P.~Zhu, W.~Huang, Q.~Yuan \emph{et~al.}, ``Quantitative coherence
  analysis with an x-ray talbot–lau interferometer,'' \emph{Analytical and
  Bioanalytical Chemistry}, vol. 397, no.~6, pp. 2091--2094, 2010.

\bibitem{phase24}
K.~Zhang, Y.~Hong, P.~Zhu, Q.~Yuan, W.~Huang \emph{et~al.}, ``Study of osem
  with different subsets in grating-based x-ray differential phase-contrast
  imaging,'' \emph{Analytical and Bioanalytical Chemistry}, vol. 401, no.~3,
  pp. 837--844, 2011.

\bibitem{analyticInterpo}
G.~Kim, C.~Park, D.~Lee, H.~Cho, C.~Seo, S.~Park \emph{et~al.}, ``Analytic
  computed tomography reconstruction in sparse-angular sampling using a
  sinogram-normalization interpolation method,'' \emph{Journal of the Korean
  Physical Society}, vol.~73, no.~3, pp. 361--367, 2018.

\bibitem{dpcct1}
T.~K\"ohler, B.~Brendel, and E.~Roessl, ``Iterative reconstruction for
  differential phase contrast imaging using spherically symmetric basis
  functions,'' \emph{Medical Physics}, vol.~38, no.~8, pp. 4542--4545, 2011.

\bibitem{dpcct2}
M.~Nilchian, C.~Vonesch, P.~Modregger, and M.~Unser, ``Fast iterative
  reconstruction of differential phase contrast x-ray tomograms,'' \emph{Optics
  Express}, vol.~21, no.~5, pp. 5511--5528, 2013.

\bibitem{dpcct3}
J.~Fu, R.~Tan, and L.~Chen, ``Analysis and accurate reconstruction of
  incomplete data in x-ray differential phase-contrast computed tomography,''
  \emph{Analytical \& Bioanalytical Chemistry}, vol. 406, no.~3, pp. 897--904,
  2014.

\bibitem{classification1}
A.~Krizhevsky, I.~SutsKever, and G.~E. Hinton, ``Imagenet classification with
  deep convolutional neural networks,'' \emph{Advances in neural information
  processing systems}, vol.~60, no.~2, pp. 1097--1105, 2012.

\bibitem{classification2}
P.~Sermanet, D.~Eigen, X.~Zhang, M.~Mathieu, R.~Fergus, and Y.~LeCun,
  ``Overfeat: Integrated recognition, localization and detection using
  convolutional networks.'' arXiv:1312.6229v4, feb 2014.

\bibitem{U-net}
O.~Ronneberger, P.~Fischer, and T.~Brox, ``U-net: Convolutional networks for
  biomedical image segmentation,'' in \emph{International Conference on Medical
  Image Computing and Computer-Assisted Intervention}, 2015, pp. 234--241.

\bibitem{image-denoising}
H.~C. Burger, C.~J. Schuler, and S.~Harmeling, ``Image denoising: Can plain
  neural networks compete with bm3d?'' in \emph{IEEE Conference on Computer
  Vision and Pattern Recogition}, 2012.

\bibitem{reduction1}
C.~Dong, Y.~Deng, C.~C. Loy, and X.~Tang, ``Compression artifacts reduction by
  a deep convolutional network,'' in \emph{IEEE International Conference on
  Computer Vision (ICCV)}, 2015.

\bibitem{reduction2}
J.~Guo and H.~Chao, ``Building dual-domain representations for compression
  artifacts reduction,'' in \emph{European Conference on Computer Vision},
  2016, pp. 628--644.

\bibitem{CierniakDlRecon}
R.~Cierniak, ``A new approach to image reconstruction from projections using a
  recurrent neural network,'' \emph{International Journal of Applied
  Mathematics and Computer Science}, vol.~18, no.~2, pp. 147--157, 2008.

\bibitem{DLinCT1}
Y.~Han, J.~Yoo, and J.~C. Ye, ``Deep residual learning for compressed sensing
  ct reconstruction via persistent homology analysis,'' arXiv:1611.06391v2,
  2016.

\bibitem{DLinCT2}
J.~Gu and J.~C. Ye, ``Multi-scale wavelet domain residual learning for
  limited-angle ct reconstruction,'' arXiv:1703.01382v1, 2017.

\bibitem{convolutionalforCT}
K.~H. Jin, M.~T. McCann, E.~Froustey, and M.~Unser, ``Deep convolutional neural
  network for inverse problems in imaging,'' \emph{IEEE Transactions on Image
  Processing}, vol.~26, no.~9, pp. 4509--4522, 2017.

\bibitem{PeltDlRecon}
D.~M. Pelt and J.~A. Sethian, ``A mixed-scale dense convolutional neural
  network for image analysis,'' \emph{PNAS 2017}, 2017.

\bibitem{ZhichengDLRecon}
Z.~Zhang, X.~Liang, X.~Dong, Y.~Xie, and G.~Cao, ``A sparse-view ct
  reconstruction method based on combination of densenet and deconvolution,''
  \emph{IEEE Transactions on Medical Imaging}, vol.~37, no.~6, pp. 1407--1417,
  2018.

\bibitem{denseNet}
G.~Huang, Z.~Liu, L.~van, der Maaten, and K.~Q. Weinberger, ``Densely connected
  convolutional networks,'' \emph{IEEE CVPR 2017}, 2017.

\bibitem{deconvolution}
L.~Xu, J.~S.~J. Ren, C.~Liu, and J.~Jia, ``Deep convolutional neural network
  for image deconvolution,'' \emph{Advances in neural information processing
  systems}, pp. 1790--1798, 2014.

\bibitem{sinoInterpo1}
H.~Lee, J.~Lee, and S.~Cho, ``View-interpolation of sparsely sampled sinogram
  using convolutional neural network,'' in \emph{Proceedings of SPIE}, vol.
  10133, 2017.

\bibitem{residuallearning}
K.~He, X.~Zhang, S.~Ren, and J.~Sun, ``Deep residual learning for image
  recognition,'' in \emph{IEEE Conference on Computer Vision and Pattern
  Recognition}, 2016, pp. 770--778.

\bibitem{sinoInterpo2}
H.~Lee, J.~Lee, H.~Kim, B.~Cho, and S.~Cho, ``Deep-neural-network based
  sinogram synthesis for sparse-view ct image reconstruction,'' \emph{IEEE
  Transactions on Radiation and Plasma Medical Science}, p.~1, 2018.

\bibitem{sinoInterpo3}
D.~Lee, S.~Choi, and H.-J. Kim, ``High quality imaging from sparsely sampled
  computed tomography data with deep learning and wavelet transform in various
  domains,'' \emph{Medical Physics}, 2018.

\bibitem{deepDepth}
G.~Huang, Y.~Sun, Z.~Liu, D.~Sedra, and K.~Weinberger, ``Deep networks with
  stochastic depth,'' in \emph{In ECCV}, 2016.

\bibitem{relu}
V.~Nair and G.~E. Hinton, ``Rectified linear units improve restricted boltzmann
  machines,'' in \emph{In Proceedings of 27th International Conference on
  Machine Learning}, 2010, pp. 807--814.

\bibitem{batchnorm}
S.~Ioffe and C.~Szegedy, ``Batch normalization: Accelerating deep network
  training by reducing internal covariate shift,'' arXiv:1502.03167v3, 2015.

\bibitem{deepSupervision}
C.-Y. Lee, S.~Xie, P.~Gallagher, and Z.~Zhang, ``Deeply-supervised nets,'' in
  \emph{In AISTATS}, 2015.

\bibitem{ms_ssim}
H.~Zhao, O.~Gallo, I.~Frosio, and J.~Kautz, ``Loss functions for image
  restoration with neural networks,'' \emph{IEEE Transactions on Computational
  Imaging}, vol.~3, pp. 47--57.

\bibitem{adaboundOp}
L.~Luo, Y.~Xiong, Y.~Liu, and X.~Sun, ``Adaptive gradient methods with dynamic
  bound of learning rate,'' in \emph{ICIR 2019}, 2019.

\bibitem{dataCitation}
H.~R. Roth, L.~Lu, A.~Seff, K.~M. Cherry, J.~Hoffman, S.~Wang, J.~Liu,
  E.~Turbey, and R.~M. Summers, ``A new 2.5d representation for lymph node
  detection in ct,'' The Cancer Imaging Archive.
  \url{http://doi.org/10.7937/K9/TCIA.2015.AQIIDCNM}, 2015.

\bibitem{publicationCitation1}
H.~R. Roth, L.~Lu, A.~Seff, K.~M. Cherry, J.~Hoffman, S.~Wang \emph{et~al.},
  ``A new 2.5d representation for lymph node detection using random sets of
  deep convolutional neural network observations.'' in \emph{Medical Image
  Computing and Computer-Assisted Intervention-MICCAI 2014}, 2014, pp.
  520--527.

\bibitem{publicationCitation2}
A.~Seff, L.~Lu, K.~M. Cherry, H.~Roth, J.~Liu, S.~Wang, J.~Hoffman, E.~B.
  Turbey, and R.~M. Summers, ``2d view aggregation for lymph node detection
  using a shallow hierarchy of linear classifiers.'' in \emph{Medical Image
  Computing and Computer-Assisted Intervention-MICCAI 2014}, 2014, pp.
  544--552.

\bibitem{tciaCitation}
K.~Clark, B.~Vendt, K.~Smith, J.~Freymann, J.~Kirby, P.~Koppel, S.~Moore,
  S.~Phillips, D.~Maffitt, M.~Pringle, L.~Tarbox, and F.~Prior, ``The cancer
  imaging archive (tcia): Maintaining and operating a public information
  repository,'' \emph{Journal of Digital Imaging}, vol.~26, no.~6, pp.
  1045--1057, December 2013.

\bibitem{adam}
D.~P. Kingma and J.~L. Ba, ``Adam: A method for stochastic optimization,'' in
  \emph{ICIR 2015}, 2015.

\bibitem{fullReference}
L.~Zhang, L.~Zhang, X.~Mou, and D.~Zhang, ``A comprehensive evaluation of full
  reference image quality assessment algorithms,'' in \emph{In Proceedings of
  International Conference on Image Processing}, 2012, pp. 1477--1480.

\bibitem{fsim}
------, ``Fsim: A feature similarity index for image quality assessment,''
  \emph{IEEE Transactions on Image Processing}, vol.~20, no.~8, pp. 2378--2386,
  2011.

\bibitem{ssim}
Z.~Wang, A.~C. Bovik, H.~R. Sheikh, and E.~P. Simoncelli, ``Image quality
  assessment: From error visibility to structural similarity,'' \emph{IEEE
  Transactions on Image Processing}, vol.~13, no.~4, pp. 600--612, 2004.

\bibitem{iw_ssim}
Z.~Wang and Q.~Li, ``Information content weighting for perceptual image quality
  assessment,'' \emph{IEEE Transactions on Image processing}, vol.~20, no.~5,
  pp. 1185--1198, 2011.

\end{thebibliography}

\end{document}